\documentclass[oneside,reqno]{amsart}
\usepackage{beraserif}
\usepackage{berasans}
\usepackage[T1]{fontenc}
\usepackage[latin9]{inputenc}
\usepackage{color}
\usepackage[english]{babel}
\usepackage{array}
\usepackage{prettyref}
\usepackage{float}
\usepackage{units}
\usepackage{multirow}
\usepackage{amsbsy}
\usepackage{amstext}
\usepackage{amsthm}
\usepackage{amssymb}
\usepackage{stmaryrd}
\usepackage{setspace}
\usepackage{xargs}[2008/03/08]
\usepackage{microtype}
\usepackage[unicode=true,pdfusetitle,
 bookmarks=true,bookmarksnumbered=false,bookmarksopen=false,
 breaklinks=false,pdfborder={0 0 1},backref=false,colorlinks=true]
 {hyperref}
\hypersetup{
 citecolor=blue,linkcolor=magenta,anchorcolor=green}

\makeatletter

\providecommand{\tabularnewline}{\\}

\numberwithin{equation}{section}
\usepackage{changebar}
\providecommand{\lyxadded}[3]{}

\renewcommand{\lyxadded}[3]{
  {\protect\cbstart\color{lyxadded}{}#3\protect\cbend}
}

\@ifundefined{date}{}{\date{}}
\usepackage{upgreek}

\makeatother

\begin{document}
\selectlanguage{english}
\global\long\def\set#1#2{\left\{  #1\, |\, #2\right\}  }%

\global\long\def\cyc#1{\mathbb{Q}\left[\zeta_{#1}\right]}%

\global\long\def\mat#1#2#3#4{\left(\begin{array}{cc}
 #1  &  #2\\
 #3  &  #4 
\end{array}\right)}%

\global\long\def\Mod#1#2#3{#1\equiv#2\, \left(\mathrm{mod}\, \, #3\right)}%

\global\long\def\inv{^{\,\textrm{-}1}}%

\global\long\def\map#1#2#3{#1\!:\!#2\!\rightarrow\!#3}%

\global\long\def\Map#1#2#3#4#5{\begin{split}#1:#2  &  \rightarrow#3\\
 #4  &  \mapsto#5 
\end{split}
 }%

\global\long\def\fact#1#2{#1\slash#2}%

\global\long\def\gl#1#2{\mathsf{GL}_{#2}\!\left(#1\right)}%

\global\long\def\SL{\mathrm{SL}_{2}\!\left(\mathbb{Z}\right)}%

\global\long\def\zn#1{\left(\mathbb{Z}/\!#1\mathbb{Z}\right)^{\times}}%

\global\long\def\sn#1{\mathbb{S}_{#1}}%

\global\long\def\aut#1{\mathrm{Aut\mathit{\left(#1\right)}}}%

\global\long\def\FA#1{\vert#1\vert}%

\global\long\def\FB#1{\mathtt{Z}^{2}\!(#1)}%

\global\long\def\FC#1{#1^{{\scriptscriptstyle \flat}}}%

\global\long\def\FD#1{#1^{{\scriptscriptstyle \times}}}%

\global\long\def\FE#1{\mathtt{x}_{#1}}%

\global\long\def\FF#1#2{\mathrm{Fix}_{#1}\left(#2\right)}%

\global\long\def\FI#1{#1_{{\scriptscriptstyle \pm}}}%

\global\long\def\FJ#1#2#3{\uptheta\!\Bigl[\!{#1\atop #2}\!\Bigr]\!\left(#3\right)}%

\global\long\def\ex#1{\mathtt{e}^{2\mathtt{i}\pi#1}}%

\newcommandx\exi[3][usedefault, addprefix=\global, 1=, 2=]{\mathtt{e}^{\mathtt{{\scriptscriptstyle \textrm{#1}}}\frac{#2\pi\mathtt{i}}{#3}}}%

\global\long\def\tw#1#2#3{\mathfrak{#1}_{#3}^{{\scriptscriptstyle \left(\!#2\!\right)}}}%

\global\long\def\ver#1{\mathtt{Ver}_{#1}}%

\global\long\def\stab#1#2{#1_{#2}}%

\global\long\def\cft{\mathscr{C}}%

\global\long\def\ann#1{\ker\mathfrak{#1}}%

\global\long\def\cent#1#2{\mathtt{C}_{#1}\!\left(#2\right)}%

\global\long\def\ch{\boldsymbol{\uprho}}%

\global\long\def\chb#1{\xi_{#1}}%

\global\long\def\chg{\mathfrak{g}}%

\global\long\def\tgr{G}%

\global\long\def\qd#1{\mathtt{d}_{#1}}%
\global\long\def\cw#1{\mathtt{h}_{#1}}%

\global\long\def\tcl{\boldsymbol{\mathfrak{1}}}%

\global\long\def\ccl{\mathtt{C}}%

\global\long\def\om#1{\omega\!\left(#1\right)}%

\global\long\def\rami#1{\mathtt{e}_{#1}}%

\global\long\def\zm{\mathtt{Z}}%

\global\long\def\cs#1{\left\llbracket #1\right\rrbracket }%

\global\long\def\coc#1{\vartheta_{\mathfrak{#1}}}%

\global\long\def\sp#1#2{\bigl\langle#1,#2\bigr\rangle}%

\global\long\def\irr#1{\mathtt{Irr}\!\left(#1\right)}%

\global\long\def\stb#1{\mathtt{I}{}_{#1}}%

\global\long\def\sqf#1{#1^{\circ}}%

\global\long\def\CHG{\textrm{twister}}%

\global\long\def\to{\cs{\chg}}%

\global\long\def\v{\mathtt{{\scriptstyle 0}}}%

\global\long\def\wm{\mathbb{G}}%

\global\long\def\fm{\mathtt{N}}%

\global\long\def\spr{\textrm{spread}}%

\global\long\def\vera{\mathcal{V}}%

\global\long\def\svera#1{\widehat{#1}}%

\global\long\def\bd{\mathcal{D}_{\mathfrak{b}}}%

\global\long\def\rd#1{\boldsymbol{\updelta}_{#1}}%

\global\long\def\m#1{\mathfrak{\upmu}_{\mathfrak{#1}}}%

\global\long\def\voa{\mathcal{\mathbb{V}}}%

\global\long\def\bfm{\mathfrak{N}}%

\global\long\def\vb{\mathfrak{b}_{{\scriptscriptstyle 0}}}%

\global\long\def\bch#1#2{\boldsymbol{\upchi}_{\mathfrak{#1}}\!\left(#2\right)}%

\title{Orbifold deconstruction: a computational approach}
\author{Peter Bantay}
\curraddr{Institute for Theoretical Physics, E\" otv\" os L\' or\' and University, H-1117
Budapest, P\' azm\' any s. 1/A}
\keywords{conformal symmetry, orbifold models, modular tensor categories}
\begin{abstract}
We present a general deconstruction procedure aimed at recognizing
whether a given conformal model may be obtained as an orbifold of
another one, and to identify the twist group and the original model
in terms of some readily available characteristics. The ideas involved
are illustrated on the maximal deconstruction of the Ashkin-Teller
model $\mathtt{AT}_{16}$.
\end{abstract}

\maketitle

\section{Introduction}

Orbifold compactifications, i.e. the consideration of string propagation
on quotients of Minkowski space by some discrete group action, have
been introduced a long time ago \cite{Dixon_orbifolds1,Dixon_orbifolds2}
as a practical mean to generate string models that could be compatible
at low energies with the observed behavior of elementary particles.
While propagation of pointlike particles on such singular spaces (with
singularities corresponding to the fixed points of the group action)
could be problematic, it was argued that these difficulties would
not manifest themselves when considering one-dimensional strings.
Moreover, since these models arise from free string models by the
gauging of discrete symmetries, their analysis should be relatively
simple and have a nice group theoretic description, in contrast with
e.g. Calabi-Yau compactifications \cite{GSW}, which were not amenable
to exact description in those times. The idea was to find the relation
between the two-dimensional conformal models describing the inner
degrees of freedom of a string propagating on Minkowski space and
on its quotients by some group action. 

From a vantage point of view, one could say that orbifolding, i.e.
the gauging of discrete symmetries of a two-dimensional conformal
model, is a most important construction procedure in CFT, leading
to a host of new (rational) models from known ones\footnote{It has even been argued \cite{Moore.1990} that all rational conformal
models could be obtained as GKO coset models or orbifolds thereof.}. At the algebraic level, this amounts to the consideration of the
fixed-point subalgebra $\mathcal{\voa}^{G}$ of a Vertex Operator
Algebra $\mathcal{\voa}$, cf. \cite{Kac,Lepowsky-Li}, under some
(discrete) group $G\!<\!\aut{\mathcal{\voa}}$ of automorphisms, the
so-called twist group of the model: in this respect, orbifolding was
an essential ingredient in the construction of the famous Moonshine
module \cite{FLM1}. Unfortunately, the analysis of such models is
in general pretty hard, due on one hand to the need to include 'twisted
sectors' (twisted modules in VOA language), whose structure depends
heavily on the precise nature of the action of the twist group, and
on the other because of the difficulties associated with the so-called
fixed-point resolution procedure \cite{Fuchs1996}. For this reason,
only special varieties of orbifolds lend themselves to a general description:
toroidal ones \cite{Ginsparg1988}, when the original conformal model
consists of (compactified) free bosons, holomorphic orbifolds \cite{DV3,DW,DPR,Bantay1990a,Bantay1991},
when the original model is self-dual, and finally permutation orbifolds
\cite{Klemm-Schmidt,Borisov-Halpern-Schweigert,Bantay1998a,Bantay2002,Barron-Dong-Mason},
when the twist group acts by permuting identical copies of one and
the same conformal model. For more generic cases only some basic properties
of the orbifold construction are known, usually insufficient to identify
unambiguously the resulting model, so \emph{ad hoc} techniques are
required to fill in the details \cite{Dixon_orbifoldCFT,Hamidi-Vafa}.

The aim of the present note is to go in the opposite direction: given
a conformal model, recognize whether it is an orbifold of another
one, and if so, identify this original model and the corresponding
twist group \cite{sigma_dekonst}. Of course, the effectiveness of
such a deconstruction procedure depends heavily on the amount of knowledge
needed to characterize the different models. As we shall see, very
limited information is needed for deconstruction: the fusion rules
and chiral characters of the primaries \cite{DiFrancesco-Mathieu-Senechal}
usually suffice. That such a procedure could exist should not come
as a big surprise, for simple current extensions \cite{Fuchs1996a,Bantay1998}
are nothing but the deconstructions of abelian orbifolds, but the
exact details of their generalization to a non-commutative setting
are far from being obvious. 

In the next section we'll introduce our basic objects, $\CHG$s and
their twist classes, and discuss their most important properties.
Then we move on to the study of blocks and their characteristics.
\prettyref{sec:Relation-to-orbifolding} explains the relation of
these concepts to orbifolding in general, while \prettyref{sec:The-deconstruction-algorithm}
describes the deconstruction algorithm. An explicit example of deconstruction
is presented in \prettyref{sec:example}, illustrating some subtle
points of the process. Finally, we give an outlook on questions whose
study might be worth pursuing.

It should be stressed that our approach in this note is a computational
one, focusing on the algorithmic problems related to the deconstruction
procedure. Nowhere in the text shall we provide formal proofs of our
assertions. For most part such arguments might be readily supplied
(we give hints for a few of them), but there are some statements whose
actual proof would require much more elaboration, while their truth
is evidenced by the fact that the deconstruction algorithm presented
in \prettyref{sec:The-deconstruction-algorithm} leads to the expected
results whenever those can be obtained by alternate means, e.g. for
holomorphic or permutation orbifolds, while providing at the same
time a natural non-commutative generalization of (integer spin) simple
current extensions, shedding new light on some delicate aspects of
the latter. We have no doubt that the whole subject could be described
elegantly in more abstract terms along the lines of \cite{Kirillov2004,FFRS},
but the relevant techniques seem (at least to us) less amenable to
practical computations.

\section{Twisters and twist classes\label{sec:Twist-classes}}

Let's consider a rational unitary conformal model \cite{DiFrancesco-Mathieu-Senechal}.
We'll denote by $\qd p$ and $\mathtt{h}_{p}$ the quantum dimension
and conformal weight of a primary $p$, by $\chi_{p}\!\left(\tau\right)$
its chiral character, and by $\fm\!\left(p\right)$ the associated
fusion matrix, whose matrix elements are given by the fusion rules
\begin{equation}
\left[\fm\!\left(p\right)\right]_{q}^{r}=N_{pq}^{r}\label{eq:fmdef}
\end{equation}
Note that, since
\begin{equation}
\fm\!\left(p\right)\fm\!\left(q\right)=\sum_{r}N_{pq}^{r}\fm\!\left(r\right)\label{ver1}
\end{equation}
the fusion matrices generate a commutative matrix algebra over $\mathbb{C}$
(the Verlinde algebra), whose irreducible representations, all of
dimension $1$, are in one-to-one correspondence, according to Verlinde's
formula \cite{Verlinde1988}, with the primaries: to each primary
$q$ corresponds an irrep $\ch_{q}$ which assigns to the generator
$\fm\!\left(p\right)$ the complex value
\begin{equation}
\ch_{q}\!\left(p\right)=\frac{S_{pq}}{S_{\v q}}\label{chardef}
\end{equation}
where $S$ denotes the modular $S$ matrix of the model, and $\v$
labels its vacuum. In particular, one has $\qd p\!=\!\ch_{\v}\!\left(p\right)$
and
\begin{equation}
\sum_{r}N_{pq}^{r}\ch_{w}\!\left(r\right)=\ch_{w}\!\left(p\right)\ch_{w}\!\left(q\right)\label{verlinde}
\end{equation}

We call a set $\chg$ of primaries a \emph{$\mathcal{\CHG}$} if it
contains the vacuum $\v$, if all of its elements have integer conformal
weight and quantum dimension ($\cw{\alpha},\qd{\alpha}\!\in\!\mathbb{Z}$
for all $\alpha\!\in\!\chg$), and if $\chg$ is closed under fusion,
i.e. $\alpha,\beta\!\in\!\chg$ and $N_{\alpha\beta}^{p}\!>\!0$ implies
$p\!\in\!\chg$; in other words, the fusion matrices $\fm\!\left(\alpha\right)$
for $\alpha\!\in\!\chg$ generate a subring $\svera{\chg}$ of the
fusion ring. Note that, taking into account the positivity of quantum
dimensions, this last requirement amounts to the equality
\begin{equation}
\sum_{\gamma\in\chg}N_{\alpha\beta}^{\gamma}\qd{\gamma}=\qd{\alpha}\qd{\beta}\label{eq:chgcrit}
\end{equation}
The \emph{$\spr$} of the $\CHG$ $\chg$ is the (positive) rational
integer
\begin{equation}
\to=\sum_{\alpha\in\chg}\qd{\alpha}^{2}\label{spreaddef}
\end{equation}
A $\CHG$ is abelian if $\qd{\alpha}\!=\!1$ for all $\alpha\!\in\!\chg$.
In particular, the vacuum primary $\v$ forms in itself an abelian
$\CHG$, the \emph{trivial $\CHG$}. Abelian $\CHG$s are, in the
traditional language of CFT, nothing but groups of integer spin simple
currents. 

A \emph{twist class} $\ccl$ of a $\CHG$ $\chg$ is a (maximal) set
of primaries such that, for $p\!\in\!\ccl$, all the representations
$\ch_{p}$ of the Verlinde algebra restrict to one and the same representation
$\ch_{\ccl}$ of the subalgebra $\svera{\chg}$ generated by the twister.
We shall denote by $\alpha\!\left(\ccl\right)$ the value assigned
to $\fm\!\left(\alpha\right)$ by this representation $\ch_{\ccl}$,
so that $p\!\in\!\ccl$ precisely when $S_{\alpha p}\!=\!\alpha\!\left(\ccl\right)S_{\v p}$
for all $\alpha\!\in\!\chg$. It is immediate that twist classes partition
the set of primaries, and that their number equals the cardinality
$\FA{\chg}$ of the twister. 

The \emph{extent} of a twist class $\ccl$ is the quantity
\begin{equation}
\cs{\ccl}=\frac{1}{\sum_{p\in\ccl}S_{\v p}^{2}}\label{eq:tclext}
\end{equation}
which may be shown the be a (positive) rational integer dividing the
$\spr$ $\to\!=\!\sum_{\alpha\in\mathfrak{g}}\qd{\alpha}^{2}$ of
the $\CHG$. Using Eq.\eqref{verlinde}, one may derive the orthogonality
relations (with the bar denoting complex conjugation)
\begin{equation}
\sum_{\alpha\in\chg}\alpha\!\left(\ccl_{1}\right)\overline{\alpha\!\left(\ccl_{2}\right)}=\cs{\ccl_{1}}\delta_{\ccl_{1}\ccl_{2}}\label{eq:orth2}
\end{equation}
and
\begin{equation}
\sum_{\ccl}\frac{\alpha\!\left(\ccl\right)\overline{\beta}\!\left(\ccl\right)}{\cs{\ccl}}=\delta_{\alpha\beta}\label{eq:orth1}
\end{equation}
for $\alpha,\beta\!\in\!\chg$, where the last sum runs over all twist
classes. It follows from Eq.\eqref{eq:orth2} that
\begin{equation}
\sum_{\alpha\in\chg}\FA{\alpha\!\left(\ccl\right)}^{2}=\cs{\ccl}\label{eq:tclext2}
\end{equation}

The twist class containing the vacuum $\v$ is the \emph{trivial class}
$\tcl$: note that $\alpha\!\left(\tcl\right)\!=\!\qd{\alpha}$ by
the above definition. Using the modular relation and Eq.\eqref{verlinde},
one may show that all elements of the $\CHG$ belong to the trivial
class, i.e. $\chg\!\subseteq\!\tcl$. Obviously, the extent of the
trivial class equals the $\spr$ of the twister, while its size is
given by
\begin{equation}
\FA{\tcl}=\frac{1}{\to}\sum_{\alpha\in\chg}\qd{\alpha}\mathrm{Tr}\left(\fm\!\left(\alpha\right)\right)\label{eq:trivtclsize}
\end{equation}
A most important property of the trivial twist class that follows
ultimately from Eq.\eqref{eq:orth2} is the \emph{product rule}: if
$p$ belongs to the trivial twist class and $N_{pq}^{r}\!>\!0$, then
$q$ and $r$ belong necessarily to the same twist class.

For an integer $n$ and a twist class $\ccl$, there is a unique twist
class $\ccl^{n}$, the $n$\emph{-th power} of $\ccl$, for which
\begin{equation}
\alpha\!\left(\ccl^{n}\right)=\cs{\ccl}\sum_{p,q\in\ccl}N_{\alpha p}^{q}S_{0p}S_{0q}\mathtt{e}{}^{2\pi\mathtt{i}n\left(\mathtt{h}_{p}-\cw q\right)}\label{eq:powerdef}
\end{equation}
for all $\alpha\!\in\!\chg$. The order of a twist class $\ccl$ is
the smallest positive integer $n$ such that $\ccl^{n}$ equals the
trivial class, i.e. $\alpha\!\left(\ccl^{n}\right)\!=\!\alpha\!\left(\tcl\right)\!=\!\qd{\alpha}$
for all $\alpha\!\in\!\chg$. We note that the order of a twist class
may be shown to always divide its extent. 

Since a $\CHG$ consists of simple objects of a modular tensor category
with integer dimension, trivial twists ($\cw{\alpha}\!\in\!\mathbb{Z}$
for $\alpha\!\in\!\chg$) and is closed under fusion, it follows from
results on Tannakian categories \cite{Deligne1990} that there exists
a finite group $\tgr$ of order $\FA G\!=\!\sum_{\alpha\in\mathfrak{g}}\qd{\alpha}^{2}\!=\!\to$
whose representation ring coincides with the fusion subring $\svera{\chg}$
generated by the $\CHG$. In particular, to each element $\alpha\!\in\!\chg$
there corresponds an irreducible character $\FC{\alpha}\!\in\!\irr{\tgr}$
of degree $\FC{\alpha}\!\left(1\right)\!=\!\qd{\alpha}$, and these
satisfy the multiplication rule
\begin{equation}
\FC{\alpha}\FC{\beta}=\sum_{\gamma\in\chg}N_{\alpha\beta}^{\gamma}\FC{\gamma}\label{eq:chgmult}
\end{equation}
from which one can infer \cite{Lux-Pahlings} the values of the characters
$\FC{\alpha}\!\in\!\irr{\tgr}$ on the different conjugacy classes
of $G$: to each twist class $\ccl$ corresponds a conjugacy class
$\FC{\ccl}$ of $\tgr$, with the trivial twist class $\tcl$ corresponding
to the the trivial conjugacy class containing solely the identity
element, and this correspondence is such that
\begin{equation}
\alpha\!\left(\ccl\right)=\FC{\alpha}\!\left(\FC{\ccl}\right)\label{eq:cclchar}
\end{equation}
This implies, as a consequence of the second orthogonality relations
for group characters, that the size of the conjugacy $\FC{\ccl}$
equals

\begin{equation}
\FA{\FC{\ccl}}=\frac{\to}{\cs{\ccl}}\label{eq:cclsize}
\end{equation}

While the knowledge of the representation ring does determine many
properties of the twist group, in particular its character table and
normal structure, it does not determine its isomorphism type uniquely.
A famous example of this phenomenon is that of the groups $\mathbb{D}_{8}$
(the dihedral group of order $8$, i.e. the symmetry group of a square)
and the group $\mathsf{Q}$ of unit quaternions, which have identical
representation rings but are nevertheless not isomorphic \cite{Lux-Pahlings}.
But in our case one can pin down uniquely the twist group associated
to the $\CHG$ by exploiting the underlying braided monoidal structure,
which makes the subring $\svera{\chg}$ a $\lambda$-ring \cite{M.F.Atiyah1969},
and this extra structure should match that of the representation ring
of $G$. Indeed, Eq.\eqref{eq:powerdef}, which may be deduced along
the lines of \cite{Bantay1997a,Giorgetti2017} by considering traces
of suitable braidings, allows one to define the Adams operation $\Psi^{n}$
on $\irr G$ via the rule
\begin{equation}
\left(\Psi^{n}\FC{\alpha}\right)\!\left(\FC{\ccl}\right)=\alpha\!\left(\ccl^{n}\right)\label{eq:Adamsdef}
\end{equation}
and this extra information is usually sufficient to determine $G$
up to isomorphism \cite{Isaacs,Lux-Pahlings}. Notice that Eqs.\eqref{eq:powerdef}
and \eqref{eq:Adamsdef} imply that the (higher) Frobenius-Schur indicators
\cite{Bantay1997a,Siu-HungNg2010} of the primaries $\alpha\!\in\!\chg$
agree with those of the corresponding characters $\FC{\alpha}\!\in\!\irr{\tgr}$.

To sum up, to any $\CHG$ $\chg$ is associated a group $G$ whose
representation ring is isomorphic with $\svera{\chg}$ as a $\lambda$-ring,
and in particular the elements of $\chg$ correspond to irreps of
$G$, while the twist classes of $\chg$ to conjugacy classes of $G$.
As we shall explain in \prettyref{sec:Relation-to-orbifolding}, each
$G$-orbifold contains a set of primaries that form a $\CHG$ such
that the associated group is isomorphic with $G$. Orbifold deconstruction
is the process of identifying a conformal model whose $G$-orbifold
is the conformal model we started with from information related solely
to the orbifold and the corresponding $\CHG$.

\section{Blocks\label{sec:Blocks}}

As we have seen in the previous section, a $\CHG$ partitions the
primaries of a conformal model into twist classes which are in one-to-one
correspondence with the conjugacy classes of the twist group. It turns
out that a finer partition plays a major role in orbifold deconstruction,
the partition of the primaries into \emph{blocks}.

The primaries $p$ and $q$ belong to the same block with respect
to the $\CHG$ $\chg$ if there exists some $\alpha\!\in\!\chg$ such
that $N_{\alpha p}^{q}\!>\!0$. Put differently, this means that $\wm_{pq}\!>\!0$
for the non-negative integer matrix
\begin{equation}
\wm=\sum_{\alpha\in\chg}\qd{\alpha}\fm\!\left(\alpha\right)\label{eq:wmdef}
\end{equation}
In particular, the elements of the $\CHG$ themselves form a block,
the \emph{vacuum block} $\vb$. It is straightforward that the blocks
partition the set of primaries, and it follows from the product rule
and the containment $\chg\!\subseteq\!\tcl$ that two primaries that
belong to the same block also belong to the same twist class. Consequently,
each twist class is actually a disjoint union of blocks, with the
vacuum block contained in the trivial class.

Note that, according to the above definition, the fusion matrices
$\fm\!\left(\alpha\right)$ for $\alpha\!\in\!\chg$ can be simultaneously
brought into a block-diagonal form after a suitable rearrangement
of the primaries:
\begin{equation}
\fm\!\left(\alpha\right)=\bigoplus_{\mathfrak{b}}\fm_{\mathfrak{b}}\!\left(\alpha\right)\label{eq:blockdec}
\end{equation}
where $\mathfrak{b}$ runs over the blocks of $\chg$, and in particular
$\wm=\bigoplus_{\mathfrak{b}}\wm_{\mathfrak{b}}$ with each $\wm_{\mathfrak{b}}$
a positive integer matrix. This implies that the blocks correspond
to integral representations of $\svera{\mathfrak{\chg}}$, since the
corresponding fusion matrices $\fm_{\mathfrak{b}}\!\left(\alpha\right)$
have non-negative integer entries. On the other hand, we know that
the irreducible complex representations of $\svera{\chg}$ are among
the $\ch_{\ccl}$, hence the integral representation corresponding
to a block $\mathfrak{b}$ decomposes into a direct sum of the latter,
with the multiplicity of $\ch_{\ccl}$ given by the \emph{overlap}
\begin{equation}
\sp{\mathfrak{b}}{\ccl}=\frac{1}{\to}\sum_{\alpha\in\chg}\overline{\alpha\!\left(\ccl\right)}\mathrm{Tr~}\fm_{\mathfrak{b}}\!\left(\alpha\right)=\sum_{p\in\ccl}\sum_{q\in\mathfrak{b}}\FA{S_{pq}}^{2}\label{overlapdef}
\end{equation}
of the block $\mathfrak{b}$ and the twist class $\ccl$, which is
always a non-negative rational integer according to the above. As
a consequence of the symmetry of the matrix $S$, the overlap satisfies
the reciprocity relation
\begin{equation}
\sum_{\mathfrak{b}\subseteq\ccl_{1}}\sp{\mathfrak{b}}{\ccl_{2}}=\sum_{\mathfrak{b}\subseteq\ccl_{2}}\sp{\mathfrak{b}}{\ccl_{1}}\label{eq:reciprocity}
\end{equation}

\noindent for any two twist classes $\ccl_{1}$ and $\ccl_{2}$. 

A consequence of Eq.\eqref{eq:tclext2} is that the overlap of the
vacuum block $\vb$ with any twist class is $1$. A similar result
follows from Perron's theorem \cite{Gantmacher1959} applied to the
positive matrix $\wm_{\mathfrak{b}}$: the overlap of any block $\mathfrak{b}$
with the trivial twist class $\tcl$ is equal to $1$
\begin{equation}
\sp{\mathfrak{b}}{\tcl}=1\label{eq:trivoverlap}
\end{equation}

Exploiting the unitarity of $S$, summing Eq.\eqref{overlapdef} over
all blocks gives
\begin{equation}
\FA{\ccl}=\sum_{\mathfrak{b}}\sp{\mathfrak{b}}{\ccl}\label{eq:cclsize2}
\end{equation}
for the number of primaries in the twist class $\ccl$, while summing
over twist classes leads to
\begin{equation}
\FA{\mathfrak{b}}=\sum_{\ccl}\sp{\mathfrak{b}}{\ccl}\label{eq:blocksize}
\end{equation}
for the cardinality of the block $\mathfrak{b}$. 

Combining Eq.\eqref{eq:trivoverlap} with Eq.\eqref{eq:reciprocity},
one gets that the number of different blocks contained in a given
twist class can be expressed as
\begin{equation}
\#\set{\mathfrak{b}}{\mathfrak{b\!\subseteq\!\ccl}}=\sum_{\mathfrak{b}\subseteq\ccl}\sp{\mathfrak{b}}{\tcl}=\sum_{\mathfrak{b}\subseteq\tcl}\sp{\mathfrak{b}}{\ccl}\label{eq:blocknum}
\end{equation}
i.e. it equals the sum of overlaps of the twist class with all blocks
contained in the trivial class. Summing Eq.\eqref{eq:blocknum} over
all twist classes and taking into account Eq.\eqref{eq:blocksize}
gives
\begin{equation}
\sum_{\ccl}\#\set{\mathfrak{b}}{\mathfrak{b\!\subseteq\!\ccl}}=\sum_{\ccl}\sum_{\mathfrak{b}\subseteq\tcl}\sp{\mathfrak{b}}{\ccl}=\sum_{\mathfrak{b}\subseteq\tcl}\FA{\mathfrak{b}}=\FA{\tcl}\label{eq:blockno}
\end{equation}
i.e. the total number of blocks with respect to $\chg$ is equal to
the size (cardinality) of the trivial twist class.

Using Perron's theorem \cite{Gantmacher1959} for the matrix $\wm_{\mathfrak{b}}$,
combined with some elementary Galois theory, one can show that the
ratio of quantum dimensions inside a block $\mathfrak{b}$ are always
rational numbers, hence there exists a largest algebraic integer $\bd$
such that $\qd p\!\in\!\bd\mathbb{Z}_{{\scriptscriptstyle +}}$ for
all $p\!\in\!\mathfrak{b}$. As a consequence, the ratio
\begin{equation}
\m b\!=\!\frac{1}{\bd^{2}}\sum\limits _{p\in\mathfrak{b}}\qd p^{2}\label{eq:reldimsdef}
\end{equation}
is a rational integer,  which may be shown to divide the extent of
the twist class containing $\mathfrak{b}$. We note that $\bd$ equals
in most cases the minimum $\min\!\set{\qd p}{p\!\in\!\mathfrak{b}}$
of the quantum dimensions of primaries from $\mathfrak{b}$.

The above results have a nice representation theoretic interpretation.
To each block $\mathfrak{b}$ is associated a subgroup (more precisely,
a conjugacy class of subgroups) of the twist group $G$ - the \emph{inertia
group} $\stb{\mathfrak{b}}$ of the block - and a 2-cocycle $\coc b\!\in\!\FB{\stb{\mathfrak{b}}}$.
Denoting by $\rami{\mathfrak{b}}$ the multiplicative order of the
cohomology class of $\coc b$, there is a one-to-one correspondence
$p\!\leftrightarrow\!\xi_{p}$ between primaries $p\!\in\!\mathfrak{b}$
and (projective) irreducible characters $\chb p\!\in\!\irr{\stb{\mathfrak{b}}\!\mid\!\coc b}$
of the inertia subgroup $\stb{\mathfrak{b}}$ with cocycle $\coc b$,
such that
\begin{equation}
\chb p\!\left(1\right)=\rami{\mathfrak{b}}\frac{\qd p}{\bd}\label{eq:inertdims}
\end{equation}
and
\begin{equation}
\sum_{q\in\mathfrak{b}}N_{\alpha p}^{q}\chb q=\FC{\alpha}_{\mathfrak{b}}\chb p\label{eq:inertmult}
\end{equation}
for all $\alpha\!\in\!\chg$, where $\FC{\alpha}_{\mathfrak{b}}$
denotes the restriction to $\stb{\mathfrak{b}}$ of the irreducible
character $\FC{\alpha}$ of the twist group $G$ associated to the
primary $\alpha\!\in\!\chg$. It follows from \eqref{eq:inertdims}
and Burnside's theorem \cite{Isaacs,Lux-Pahlings} that the order
of the inertia subgroup is given by
\begin{equation}
\FA{\stb{\mathfrak{b}}}=\sum_{p\in\mathfrak{b}}\chb p\!\left(1\right)^{2}=\rami{\mathfrak{b}}^{2}\m b\label{eq:inertsize}
\end{equation}
while the overlap $\sp{\mathfrak{b}}{\ccl}$ counts the number of
$\coc b$-regular conjugacy classes\footnote{Recall that an element $x\!\in\!\stb{\mathfrak{b}}$ belongs to a
$\coc b$-regular conjugacy class if $\coc{\mathfrak{b}}\!\left(x,y\right)\!=\!\coc b\!\left(y,x\right)$
whenever $y\!\in\!\stb{\mathfrak{b}}$ commutes with $x$, cf. \cite{Isaacs,Lux-Pahlings}.} of $\stb{\mathfrak{b}}$ meeting the conjugacy class $\FC{\ccl}$
of $G$.

To each block $\mathfrak{b}$ is associated the non-negative matrix
\begin{equation}
\fm\!\left(\mathfrak{b}\right)=\frac{1}{\rami{\mathfrak{b}}\m b}\sum_{p\in\mathfrak{b}}\frac{\qd p}{\bd}\fm\!\left(p\right)\label{eq:bfmdef}
\end{equation}
These matrices\emph{ }may be shown to form a ring, i.e. the product
of any two of them may be expressed as a sum
\begin{equation}
\fm\!\left(\mathfrak{a}\right)\fm\!\left(\mathfrak{b}\right)=\sum_{\mathfrak{c}}\mathfrak{\bfm_{ab}^{c}}\fm\!\left(\mathfrak{c}\right)\label{eq:bfmprod}
\end{equation}
with integer \emph{block-fusion} \emph{coefficients} given by
\begin{equation}
\mathfrak{\bfm_{ab}^{c}}=\frac{\rami{\mathfrak{c}}}{\rami{\mathfrak{a}}\m a\rami{\mathfrak{b}}\m b}\sum_{p\in\mathfrak{a}}\sum_{q\in\mathfrak{b}}\sum_{r\in\mathfrak{c}}N_{pq}^{r}\frac{\qd p}{\mathcal{D}_{\mathfrak{a}}}\frac{\qd q}{\bd}\frac{\qd r}{\mathcal{D}_{\mathfrak{c}}}\label{bfmelms}
\end{equation}

If the block $\mathfrak{b}$ is contained in a twist class $\ccl$
of order $n$, i.e. such that $\alpha\!\left(\ccl^{n}\right)\!=\!\alpha\!\left(\tcl\right)\!=\!\qd{\alpha}$
for all $\alpha\!\in\!\chg$, then one may show, using some elementary
character theory, that the conformal weights of all primaries from
$\mathfrak{b}$ differ from each other by integer multiples of $\frac{1}{n}$.
In other words, denoting by $\cw{\mathfrak{b}}\!=\!\min\set{\cw p}{p\!\in\!\mathfrak{b}}$
the minimal conformal weight inside the block $\mathfrak{b}$, one
has $n\!\left(\cw p\!-\!\cw{\mathfrak{b}}\right)\!\in\!\mathbb{Z}_{{\scriptscriptstyle +}}$
for all $p\!\in\!\mathfrak{b}$. As a consequence, the character

\begin{equation}
\bch b{\tau}=\frac{\rami{\mathfrak{b}}}{\bd}\sum_{p\in\mathfrak{b}}\qd p\chi_{p}\!\left(\tau\right)\label{eq:blockchardef}
\end{equation}
of the block $\mathfrak{b}$ will have a Puiseux expansion
\begin{equation}
\bch b{\tau}=q^{\cw{\mathfrak{b}}-\frac{c}{24}}\sum_{k=0}^{\infty}a_{k}q^{\frac{k}{n}}\label{eq:blockcharexpansion}
\end{equation}
in terms of the variable $q\!=\!\exp(2\pi\textrm{i}\tau)$, with non-negative
integer coefficients $a_{k}\!\in\!\mathbb{Z}_{{\scriptscriptstyle +}}$.
In particular, the conformal weights inside blocks contained in the
trivial twist class $\tcl$ may only differ by integers, and the corresponding
expansions will be in integral powers of $q$ (apart from an overall
factor coming from the conformal weight). Let's note that, using the
matrix introduced in Eq.\eqref{eq:wmdef}, one has the following expression
for the modulus squared of the character (where, as usual, the bar
denotes complex conjugation)
\begin{equation}
\FA{\bch b{\tau}\!}^{2}=\frac{\rami{\mathfrak{b}}^{2}\m b}{\FA{\chg}}\sum_{p,q\in\mathfrak{b}}\wm_{pq}\chi_{p}\!\left(\tau\right)\overline{\chi_{q}\!\left(\tau\right)}\label{eq:blockcharsquare}
\end{equation}

\section{Relation to orbifolding\label{sec:Relation-to-orbifolding}}

Consider a rational conformal model with associated Vertex Operator
Algebra $\mathcal{\voa}$, cf. \cite{FLM1,Kac,Lepowsky-Li}, and some
(finite) group $G\!<\!\aut{\voa}$ of automorphisms. The $G$-orbifold
has associated VOA $\mathcal{\voa}^{G}$, the fixed-point subalgebra
of $\voa$. For well behaved $\voa$ (rational, $C_{2}$-cofinite,
etc.) the representation theory of $\mathcal{\voa}^{G}$ may be reconstructed
from the knowledge of the $g$-twisted modules of $\voa$ for $g\!\in\!G$,
cf. \cite{C.Dong1998,C.Dong2000} . More precisely, there is a natural
action of $G$ on the set of all $G$-twisted modules under which
an element $h\!\in\!G$ sends a $g$-twisted module to a $hgh\inv$-twisted
module, and this leads to a partition of the set of twisted modules
into sectors labeled by the conjugacy classes of $G$, with each such
sector being itself a union of $G$-orbits. Since the twisted modules
inside a $G$-orbit are related by an automorphism of $\voa$, they
have pretty similar properties while still differing from each other,
e.g. their trace functions coincide. 

The stabilizer $\stab GM\!=\!\set{h\!\in\!G}{h\!\left(M\right)\!\cong\!M}$
of a given $g$-twisted module $M$ consists of those elements $h\!\in\!G$
for which $h\!\left(M\right)$ is isomorphic to $M$. Clearly, $\stab GM$
is a subgroup of the centralizer $\cent Gg$, and the stabilizers
of different modules belonging to the same $G$-orbit are conjugate
subgroups of $G$. By the above, there is an action of the stabilizer
$\stab GM$ on $M$, but it should be stressed that this is usually
not a linear representation, but only a projective one, with an associated
$2$-cocycle $\vartheta_{M}\!\in\!Z^{2}\!\left(G_{M},\mathbb{C}\right)$.

In particular, the untwisted sector corresponding to the trivial conjugacy
class of $G$ always contains a $G$-orbit of length $1$ that consists
solely of the vacuum. Its stabilizer subgroup is the whole of $G$,
represented linearly, hence the vacuum decomposes into sectors corresponding
to the irreducible representations of $G$. Each such sector will
correspond to a primary field of the orbifold having integral conformal
weight and quantum dimension (the later being equal to the dimension
of the corresponding representation of $G$), and with fusion rules
corresponding to tensor products of irreps of the twist group $G$.
This means that these primaries of the orbifold will form a $\CHG$
$\chg$, with associated subring $\svera{\chg}$ coinciding with the
representation ring of the twist group. Exploiting Eq.\eqref{eq:Adamsdef},
the $\CHG$ does even determine the representation ring as a $\lambda$-ring,
which is essential for the identification of the twist group $G$
from the above data.

As seen in \prettyref{sec:Twist-classes}, twist classes of $\chg$
correspond to conjugacy classes of $G$, which makes possible to associate
to each twist class $\ccl$ the set of all $g$-twisted modules of
$\voa$ with $g\!\in\!\FC{\ccl}$. In particular, to the trivial twist
class $\tcl$ will correspond the untwisted sector, i.e. the set of
all ordinary $\voa$ modules. As explained above, these sets of twisted
modules are organized into orbits under the outer action of $G$,
and to each block $\mathfrak{b}\!\subseteq\!\ccl$ corresponds such
an orbit of the twist group $G$, with the stabilizer $G_{M}$ of
any twisted module $M$ belonging to it isomorphic to the inertia
subgroup $\stb{\mathfrak{b}}$ of $\mathfrak{b}$ (note that the stabilizers
of different modules from the same orbit are conjugate, hence isomorphic
subgroups of $G$), and the $2$-cocycles $\vartheta_{M}$ and $\coc b$
belonging to the same cohomology class. Finally, the primaries $p\!\in\!\mathfrak{b}$
are in one-to-one correspondence with the irreducible projective representations
$\chb p$ (with cocycle $\coc b$) of the inertia subgroup. We note
that the above argument leads to the expression
\begin{equation}
\sum_{\mathfrak{b}\subseteq\ccl}\left[G\!:\!\stb{\mathfrak{b}}\right]=\sum_{\mathfrak{b}\subseteq\ccl}\frac{\FA{\chg}}{\rami{\mathfrak{b}}^{2}\m b}\label{eq:numtwistmod}
\end{equation}
for the number of different $g$-twisted modules with $g\!\in\!\FC{\ccl}$,
and in particular
\begin{equation}
\sum_{\mathfrak{b}\subseteq\tcl}\frac{\FA{\chg}}{\rami{\mathfrak{b}}^{2}\m b}\label{eq:numdeconstprims}
\end{equation}
for the number of different $\voa$ modules.

As noticed before, the twisted modules belonging to the same $G$-orbit,
being related by an automorphism of $\voa$, have very similar properties.
In particular, the (quantum) dimension $\qd M$ of any twisted module
$M$ belonging to the orbit corresponding to the block $\mathfrak{b}$
equals the dimension
\begin{equation}
\qd{\mathfrak{b}}=\frac{\rami{\mathfrak{b}}\m b\bd}{\FA{\chg}}\geq1\label{eq:blockdimdef}
\end{equation}
of the block, its Frobenius-Schur indicator \cite{Bantay1997a} is
given by
\begin{equation}
\nu_{\mathfrak{b}}=\rami{\mathfrak{b}}\sum_{p\in\mathfrak{b}}\frac{\qd p}{\bd}\sum_{q,r\in\tcl}N_{pq}^{r}S_{\v q}S_{\v r}\mathtt{e}{}^{4\pi\mathtt{i}\left(\mathtt{h}_{r}-\cw q\right)}\label{eq:FSblock}
\end{equation}
while its trace function equals the block's character
\begin{equation}
Z_{M}\!\left(\tau\right)=\mathrm{Tr}_{M}\!\left\{ \mathtt{e}^{2\pi\mathtt{i}\tau\left(L_{0}-\nicefrac{c}{24}\right)}\right\} =\bch b{\tau}\label{eq:blockchar}
\end{equation}
and more generally, one has
\begin{equation}
\mathrm{Tr}_{M}\!\left\{ h\mathtt{e}^{2\pi\mathtt{i}\tau\left(L_{0}-\nicefrac{c}{24}\right)}\right\} =\sum_{p\in\mathfrak{b}}\chb p\!\left(h\right)\chi_{p}\!\left(\tau\right)\label{eq:GTM}
\end{equation}
for any $h\!\in\!G_{M}$. 

Taking into account multiplicities, we get from Eqs.\eqref{eq:blockchar}
and \eqref{eq:blockcharsquare} the sum rules
\begin{align}
\sum_{M}\qd MZ_{M}\!\left(\tau\right) & =\sum_{p\in\ccl}\qd p\chi_{p}\!\left(\tau\right)\label{eq:charsumrule}\\
\sum_{M}\FA{Z_{M}\!\left(\tau\right)}^{2} & =\sum_{p,q\in\ccl}\wm_{pq}\chi_{p}\!\left(\tau\right)\overline{\chi_{q}\!\left(\tau\right)}
\end{align}
where the sum on the left hand side runs over all $g$-twisted modules
$M$ with $g\!\in\!\FC{\ccl}$. In particular, the (diagonal) modular
invariant partition function of the deconstructed model reads
\begin{equation}
Z_{\mathtt{d}}\!\left(\tau,\overline{\tau}\right)=\sum_{p,q\in\tcl}\wm_{pq}\chi_{p}\!\left(\tau\right)\overline{\chi_{q}\!\left(\tau\right)}\label{eq:partfun}
\end{equation}

\section{The deconstruction algorithm\label{sec:The-deconstruction-algorithm}}

Armed with the above, we are now ready to describe the deconstruction
procedure in detail. We start from a unitary conformal model for which
we know the fusion rules $N_{pq}^{r}$, conformal weights $\cw p$,
quantum dimensions $\qd p$ and chiral characters $\chi_{p}\!\left(\tau\right)$
of the primary fields, and we wish to identify it as a non-trivial
orbifold of some other conformal model. 

The first step is to determine the (non-trivial) $\CHG$s of the model.
Each $\CHG$ leads, according to the above, to a deconstruction with
a different twist group\footnote{Although both the twist groups and the deconstructed models might
be isomorphic, but the twist group action would be different.}. Of special interest are \emph{maximal deconstructions} for which
the $\CHG$ is maximal, i.e. not contained in any other $\CHG$, for
these lead to primitive models, i.e. models that cannot be obtained
as a non-trivial orbifold of some other model. Indeed, if a $\CHG$
$\chg_{1}$ is contained in a $\CHG$ $\chg_{2}$, then the twist
group $G_{1}$ corresponding to $\chg_{1}$ will be a normal subgroup
of the twist group $G_{2}$ corresponding to $\chg_{2}$, and the
deconstruction with respect to $\chg_{1}$ will result in a ${\displaystyle \nicefrac{G_{2}}{G_{1}}}$-orbifold
of the deconstruction with respect to $\chg_{2}$.

Once we have chosen a $\CHG$ $\chg$, the next step is to determine
the corresponding partition of the primaries into twist classes. The
knowledge of the twist classes allows to determine at once the character
table of the twist group, and by computing powers of twist classes
according to Eq.\eqref{eq:powerdef}, one can even determine the power
maps of the twist group, thus making possible the precise identification
(up to isomorphism) of the latter. 

Once the twist classes are known, the next step is to determine the
partition of the primaries into blocks. Once we know the blocks, it
is straightforward to compute some of their characteristic quantities,
like the weights $\cw{\mathfrak{b}}$, dimensions $\bd$ and characters
$\bch b{\tau}$. On the other hand, the precise determination of the
inertia groups $\stb{\mathfrak{b}}$ and 2-cocycles $\coc b$ is much
more involved. Fortunately, for most of the deconstruction process
one only needs to know the multiplicative order $\rami{\mathfrak{b}}$
of $\coc b$, and this can be determined in many cases without the
actual knowledge of $\coc b$ itself, by using some simple divisibility
properties. In particular, the product $\rami{\mathfrak{b}}^{2}\m b$,
which is equal to the order of the inertia group $\stb{\mathfrak{b}}$,
always divides the extent $\cs{\ccl}$ of the twist class containing
the block $\mathfrak{b}$, being at the same time a multiple of its
order, restricting to a large extent the possible values of $\rami{\mathfrak{b}}$.
In case this is still not enough to determine $\rami{\mathfrak{b}}$
uniquely, one can restrict further the possible values using Eq.\eqref{eq:FSblock}.

Actually, most of the above considerations are superfluous if one
is only interested in the deconstructed model proper, for in that
case it is enough to perform the above computations for the blocks
contained in the trivial class $\tcl$. To each block $\mathfrak{b}\!\subseteq\!\tcl$
there will correspond
\[
\left[G\!:\!\stb{\mathfrak{b}}\right]=\frac{\FA{\chg}}{\rami{\mathfrak{b}}^{2}\m b}
\]
different primaries of the deconstructed model, each of conformal
weight $\cw{\mathfrak{b}}$, quantum dimension $\qd{\mathfrak{b}}$
and character $\bch b{\tau}$. In many cases this is already enough
to identify uniquely the deconstructed model (remember that the central
charge does not change during orbifolding/deconstruction). If there
is still some ambiguity left, one can use the block-fusion coefficients
Eq.\eqref{bfmelms}, which characterize to some extent the fusion
rules of the deconstructed model, but for one important difference:
they do not describe the fusion of the individual primaries belonging
to the corresponding blocks, but only that of the direct sum of the
modules contained in the orbits corresponding to the relevant blocks. 

\section{A worked-out example: the Ashkin-Teller model $\mathtt{A}\mathtt{T}_{16}$\label{sec:example}}

Ashkin-Teller models are unitary conformal models of central charge
$c\!=\!1$ obtained by coupling two Ising models through their energy
densities. They may be shown to be equivalent to $\mathbb{Z}_{2}$-orbifolds
(with respect to space reflection) of the compactified boson \cite{Ginsparg1988}
at suitable radii determined by the coupling. At specific values of
the coupling corresponding to compactification radii of the form $R_{\mathtt{orb}}\!=\!\sqrt{\nicefrac{N}{2}}$
with integer $N$, the resulting $\mathtt{A}\mathtt{T}_{N}$ models
are rational, with $N\!+\!7$ primary fields \cite{Cappell_porbi}.
In particular, for $N\!=\!16$ we get a total of $23$ primaries,
whose most important properties are listed in Table \ref{tab:Primaries-of-the}
(note the unusual labeling $u_{{\scriptscriptstyle +}}$ of the vacuum).
\medskip{}

\begin{table}[h]
\begin{tabular}{|c|c|c|c|}
\hline 
label  & conformal weight & dimension & character\tabularnewline
\hline 
\hline 
$u_{{\scriptscriptstyle +}}$ & $0$ & $1$ & $\frac{1}{2\eta\!\left(\tau\right)}\theta_{3}\!\left(32\tau\right)+\sqrt{\frac{\eta}{2\theta_{2}}}\!\left(\tau\right)$\tabularnewline
\hline 
$u_{{\scriptscriptstyle -}}$ & $1$ & $1$ & $\frac{1}{2\eta\!\left(\tau\right)}\theta_{3}\!\left(32\tau\right)-\sqrt{\frac{\eta}{2\theta_{2}}}\!\left(\tau\right)$\tabularnewline
\hline 
$\FI{\phi}$ & $4$ & $1$ & $\frac{1}{2\eta\!\left(\tau\right)}\theta_{2}\!\left(32\tau\right)$\tabularnewline
\hline 
$\FI{\sigma}$ & $\frac{1}{16}$ & $4$ & $\frac{1}{2}\left\{ \sqrt{\frac{\eta}{\theta_{4}}}\!\left(\tau\right)+\sqrt{\frac{\eta}{\theta_{3}}}\!\left(\tau\right)\right\} $\tabularnewline
\hline 
$\FI{\tau}$ & $\frac{9}{16}$ & $4$ & $\frac{1}{2}\left\{ \sqrt{\frac{\eta}{\theta_{4}}}\!\left(\tau\right)-\sqrt{\frac{\eta}{\theta_{3}}}\!\left(\tau\right)\right\} $\tabularnewline
\hline 
$\chi_{k}$  & \multirow{2}{*}{$\frac{k^{2}}{64}$} & \multirow{2}{*}{$2$} & \multirow{2}{*}{$\frac{1}{\eta\!\left(\tau\right)}\FJ{\frac{k}{32}}0{32\tau}$}\tabularnewline
for $1\!\leq\!k\!\leq\!15$ &  &  & \tabularnewline
\hline 
\end{tabular}

\bigskip{}
\caption{\label{tab:Primaries-of-the}Primaries of the Ashkin-Teller model
$\mathtt{A}\mathtt{T}_{16}$}
\end{table}
\noindent Here 
\begin{equation}
\FJ ab{\tau}=\sum_{n\in\mathbb{Z}}\mathtt{e}^{\mathtt{i}\pi\tau\left(n-a\right)^{2}}\mathtt{e}^{-2\pi\mathtt{i}bn}\label{eq:thetadef}
\end{equation}
for $a,b\!\in\!\mathbb{Q}$,
\begin{equation}
\eta\!\left(\tau\right)=q^{\frac{1}{24}}\prod_{n=1}^{\infty}\left(1-q^{n}\right)\label{eq:etadef}
\end{equation}
is Dedekind's eta function (with $q\!=\!\ex{\tau}$), while

\noindent 
\begin{alignat*}{2}
\theta_{2} & =\FJ{\frac{1}{2}}0{\tau}~ & =2q^{\nicefrac{1}{8}}\prod_{n=1}^{\infty}\left(1-q^{n}\right)\left(1+q^{n}\right)^{2}\\
\theta_{3} & =\FJ 00{\tau}~ & =~\prod_{n=1}^{\infty}\left(1-q^{n}\right)\left(1+q^{n-\nicefrac{1}{2}}\right)^{2}\\
\theta_{4} & =\FJ 0{\frac{1}{2}}{\tau}~ & =~\prod_{n=1}^{\infty}\left(1-q^{n}\right)\left(1-q^{n-\nicefrac{1}{2}}\right)^{2}
\end{alignat*}
are the classical theta functions of Jacobi. 

Besides the integer spin simple currents $\FI u$ and $\FI{\phi}$,
the only primary of integer conformal weight and dimension is $\chi_{{\scriptscriptstyle 8}}$.
Because $\chi_{{\scriptscriptstyle 8}}$ is fixed by all of these
simple currents, the set $\chg\!=\!\left\{ u_{{\scriptscriptstyle +}},u_{{\scriptscriptstyle -}},\phi_{{\scriptscriptstyle +}},\phi_{{\scriptscriptstyle -}},\chi_{{\scriptscriptstyle 8}}\right\} $
is closed under fusion, hence it forms a $\CHG$ of $\spr$ $8$,
the unique maximal $\CHG$ of $\mathtt{A}\mathtt{T}_{16}$. In the
sequel, we shall investigate the maximal deconstruction of $\mathtt{A}\mathtt{T}_{16}$
with respect to $\mathfrak{g}$.

There are $5$ different twist classes for the maximal $\CHG$ $\chg$,
whose properties are summarized in Table \ref{tab:Twist-classes-of}.
\begin{flushleft}
\begin{table}[H]
\begin{centering}
\begin{tabular}{|c|c|c|c|c|}
\hline 
class & fields  & extent & order & size\tabularnewline
\hline 
\hline 
$\tcl$ & $u_{{\scriptscriptstyle +}},u_{{\scriptscriptstyle -}},\phi_{{\scriptscriptstyle +}},\phi_{{\scriptscriptstyle -}},\chi_{{\scriptscriptstyle 4}},\chi_{{\scriptscriptstyle 8}},\chi_{{\scriptscriptstyle 12}}$ & $8$ & $1$ & $7$\tabularnewline
\hline 
$\ccl_{{\scriptscriptstyle +}}$ & $\sigma_{{\scriptscriptstyle +}},\tau_{{\scriptscriptstyle +}}$ & $4$ & $2$ & $2$\tabularnewline
\hline 
$\ccl_{{\scriptscriptstyle -}}$ & $\sigma_{{\scriptscriptstyle -}},\tau_{{\scriptscriptstyle -}}$ & $4$ & $2$ & $2$\tabularnewline
\hline 
$\ccl_{\mathtt{o}}$ & $\chi_{k}$ with $k$ odd & $4$ & $4$ & $8$\tabularnewline
\hline 
$\ccl_{\mathtt{e}}$ & $\chi_{k}$ with $\Mod k24$  & $8$ & $2$ & $4$\tabularnewline
\hline 
\end{tabular}
\par\end{centering}
\medskip{}
\caption{\label{tab:Twist-classes-of}Twist classes of $\protect\chg$.}
\end{table}
From the knowledge of the twist classes we get the following character
table for the twist group:
\par\end{flushleft}

\begin{flushleft}
\medskip{}
\par\end{flushleft}

\begin{center}
\begin{tabular}{|c||c|c|c|c|c|}
\cline{2-6} 
\multicolumn{1}{c|}{} & $\FC{u_{{\scriptscriptstyle +}}}$ & $\FC{u_{{\scriptscriptstyle -}}}$ & $\FC{\phi_{{\scriptscriptstyle +}}}$ & $\FC{\phi_{{\scriptscriptstyle -}}}$ & $\FC{\chi_{{\scriptscriptstyle 8}}}$\tabularnewline
\hline 
\hline 
$\FC{\tcl}$ & $1$ & $1$ & $1$ & $1$ & $2$\tabularnewline
\hline 
$\FC{\ccl_{{\scriptscriptstyle +}}}$ & $1$ & $-1$ & $1$ & $-1$ & $0$\tabularnewline
\hline 
$\FC{\ccl_{{\scriptscriptstyle -}}}$ & $1$ & $-1$ & $-1$ & $1$ & $0$\tabularnewline
\hline 
$\FC{\ccl_{\mathtt{o}}}$ & $1$ & $1$ & $-1$ & $-1$ & $0$\tabularnewline
\hline 
$\FC{\ccl_{\mathtt{e}}}$ & $1$ & $1$ & $1$ & $1$ & $-2$\tabularnewline
\hline 
\end{tabular}
\par\end{center}

\bigskip{}

There are two non-isomorphic groups with this character table \cite{Lux-Pahlings},
hence two candidates for the twist group: the dihedral group $\mathbb{D}_{8}$
and the quaternion group $\textrm{Q}$, but the latter possibility
has $3$ different conjugacy classes of elements of order $4$, while
Eq.\eqref{eq:powerdef} allows only one such class. We conclude that
the twist group is $\mathbb{D}_{8}$.

\begin{table}[!h]
\begin{raggedright}
\begin{tabular}{|c|c|c|c|c|}
\hline 
primaries  & \negmedspace{}$\bd$\negmedspace{} & \negmedspace{}\negmedspace{}$\rami{\mathfrak{b}}$\negmedspace{}\negmedspace{} & \negmedspace{}\negmedspace{}$\cw{\mathfrak{b}}$\negmedspace{}\negmedspace{} & $\bch b{\tau}$\tabularnewline
\hline 
\hline 
\negmedspace{}$u_{{\scriptscriptstyle +}},u_{{\scriptscriptstyle -}},\phi_{{\scriptscriptstyle +}},\phi_{{\scriptscriptstyle -}},\chi_{{\scriptscriptstyle 8}}$\negmedspace{} & $1$ & $1$ & $0$ & $\frac{1}{\eta\!\left(\tau\right)}\left\{ \theta_{3}\!\left(32\tau\right)+\theta_{2}\!\left(32\tau\right)+\FJ{\frac{1}{4}}0{32\tau}\right\} $\tabularnewline
\hline 
$\chi_{{\scriptscriptstyle 4}},\chi_{{\scriptscriptstyle 12}}$ & $1$ & $2$ & $\frac{1}{4}$ & $\frac{2}{\eta\!\left(\tau\right)}\left\{ \FJ{\frac{1}{8}}0{32\tau}+\FJ{\frac{3}{8}}0{32\tau}\right\} $\tabularnewline
\hline 
$\sigma_{{\scriptscriptstyle +}},\tau_{{\scriptscriptstyle +}}$ & $4$ & $1$ & $\frac{1}{16}$ & $\sqrt{\frac{\eta}{\theta_{4}}}\!\left(\tau\right)$\tabularnewline
\hline 
$\sigma_{{\scriptscriptstyle -}},\tau_{{\scriptscriptstyle -}}$ & $4$ & $1$ & $\frac{1}{16}$ & $\sqrt{\frac{\eta}{\theta_{3}}}\!\left(\tau\right)$\tabularnewline
\hline 
$\chi_{{\scriptscriptstyle 1}},\chi_{{\scriptscriptstyle 7}},\chi_{{\scriptscriptstyle 9}},\chi_{{\scriptscriptstyle 15}}$ & $2$ & $1$ & $\frac{1}{64}$ & \negmedspace{}$\frac{1}{\eta\!\left(\tau\right)}\left\{ \FJ{\frac{1}{64}}0{32\tau}+\FJ{\frac{7}{64}}0{32\tau}+\FJ{\frac{9}{64}}0{32\tau}+\FJ{\frac{7}{64}}0{32\tau}\!\right\} $\negmedspace{}\tabularnewline
\hline 
$\chi_{{\scriptscriptstyle 3}},\chi_{{\scriptscriptstyle 5}},\chi_{{\scriptscriptstyle 11}},\chi_{{\scriptscriptstyle 13}}$ & $2$ & $1$ & $\frac{9}{64}$ & \negmedspace{}$\frac{1}{\eta\!\left(\tau\right)}\left\{ \FJ{\frac{3}{64}}0{32\tau}+\FJ{\frac{5}{64}}0{32\tau}+\FJ{\frac{11}{64}}0{32\tau}+\FJ{\frac{13}{64}}0{32\tau}\!\right\} $\negmedspace{}\tabularnewline
\hline 
$\chi_{{\scriptscriptstyle 2}},\chi_{{\scriptscriptstyle 6}},\chi_{{\scriptscriptstyle 10}},\chi_{{\scriptscriptstyle 14}}$ & $2$ & $1$ & $\frac{1}{16}$ & \negmedspace{}$\frac{1}{\eta\!\left(\tau\right)}\left\{ \FJ{\frac{2}{64}}0{32\tau}+\FJ{\frac{6}{64}}0{32\tau}+\FJ{\frac{10}{64}}0{32\tau}+\FJ{\frac{14}{64}}0{32\tau}\!\right\} $\negmedspace{}\tabularnewline
\hline 
\end{tabular}
\par\end{raggedright}
\bigskip{}
\caption{\label{tab:Blocks-of-}Blocks of $\mathtt{A}\mathtt{T}_{16}$ with
respect to the maximal $\protect\CHG$ $\protect\chg$.}
\end{table}
There are $7$ blocks with respect to $\chg$, collected in Table
\ref{tab:Blocks-of-}. Notice that $\rami{\mathfrak{b}}\!=\!2$ for
the second block, providing an explicit example where non-trivial
projective representations arise. Thanks to the simple structure of
the twist group, it is possible in this case to identify the inertia
subgroups of all the blocks from the above information without much
difficulty, but we won't actually need this information. 

Only the first two blocks are of interest for identifying the deconstructed
model, since these are the ones contained in the trivial twist class.
Since the inertia groups of these blocks both have order $\rami{\mathfrak{b}}^{2}\m b\!=\!8$,
it follows that the deconstructed model has two different primaries,
with respective characters
\begin{align*}
q^{-\frac{1}{24}}\left(1+3q+4q^{2}+7q^{3}+13q^{4}+\cdots\right)\\
\intertext{and}q^{\frac{5}{24}}\left(2+2q+6q^{2}+8q^{3}+14q^{4}+\cdots\right)
\end{align*}
This already allows to identify the deconstructed model with the $SU(2)$
WZNW model at level $1$ (or, what is the same, the free boson compactified
on a circle of radius $R\!=\!\frac{1}{\sqrt{2}}$) \cite{Ginsparg1988}. 

We conclude that the Ashkin-Teller model $\mathtt{AT}_{16}$ is a
$\mathbb{D}_{8}$-orbifold of $SU(2)_{1}$. Of course, this is a well-known
result \cite{Cappell_porbi}, but we arrived at this conclusion from
a totally new approach. We note that, besides identifying unambiguously
both the deconstructed model and the twist group (up to isomorphism),
we also get non-trivial information on the structure of some of the
twisted modules of $SU(2)_{1}$.

\section{Summary and outlook}

Orbifold deconstruction, i.e. the inverse process of orbifolding is,
as we tried to demonstrate, a well defined effective procedure to
recover from some simple data characterizing an orbifold the relevant
twist group and original conformal model. Since in this case we know
the result of orbifolding right from the start, this could be particularly
helpful in the study of general properties of orbifolds. In particular,
the deconstruction procedure provides valuable information on the
structure of twisted modules, e.g. their trace functions and tensor
products. Since the structure of $g$-twisted modules does only depend
on the conjugacy class of $g$ in the whole automorphism group of
the deconstructed model, this information pertains to the structure
of any orbifold of the deconstructed model whose twist group contains
elements conjugate to some element of $G$. 

Another interesting aspect of orbifold deconstruction is related to
modular invariants. Indeed, the (diagonal) modular invariant of the
deconstructed model, cf. Eq.\eqref{eq:partfun}, is a non-trivial
modular invariant of extension type \cite{DiFrancesco-Mathieu-Senechal}
of the orbifold, and it seems likely that many such invariants can
be related to suitable deconstructions. 

Finally, orbifold deconstruction might prove useful in attempts to
classify rational conformal models. For one thing, the classification
problem can be reduced to that of primitive models, i.e. the ones
that don't have nontrivial $\CHG$s, since all other models, being
suitable orbifolds of the primitive ones, can be classified by group
theoretic means. Besides this, primitive models can be grouped together
if they have identical orbifolds with respect to suitable twist groups,
i.e. if they arise from maximal deconstructions of one and the same
conformal model, since this points to some close relation between
them. In any case, pursuing this line of thought seems to be a worthy
undertaking.

\end{document}